\documentclass[sigconf]{acmart}

\AtBeginDocument{%
  }

\copyrightyear{2022} 
\acmYear{2022} 
\setcopyright{acmcopyright}\acmConference[KDD '22]{Proceedings of the 28th ACM SIGKDD Conference on Knowledge Discovery and Data Mining}{August 14--18, 2022}{Washington, DC, USA}
\acmBooktitle{Proceedings of the 28th ACM SIGKDD Conference on Knowledge Discovery and Data Mining (KDD '22), August 14--18, 2022, Washington, DC, USA}
\acmPrice{15.00}
\acmDOI{10.1145/3534678.3539228}
\acmISBN{978-1-4503-9385-0/22/08}

\usepackage{amsfonts}
\usepackage{dsfont}

\newcommand{\argmin}{\mathop{\rm arg~min}\limits}
\usepackage{wrapfig}
\usepackage{enumitem}
\usepackage{lipsum}
\usepackage{mathtools}
\usepackage{cleveref}
\usepackage{autonum}   %%put after hyperref, cleverref
\usepackage{graphicx,subfigure}

\crefname{equation}{Equation}{Equation}
\crefname{figure}{Figure}{Figures}
\crefname{table}{Table}{Table}
\crefname{appendix}{Appendix}{Appendix}
\crefname{section}{Section}{Section}

\setlist[itemize]{leftmargin=*}

\makeatletter
\newcommand\footnoteref[1]{\protected@xdef\@thefnmark{\ref{#1}}\@footnotemark}
\makeatother

\begin{document}

\title{Predicting Opinion Dynamics via Sociologically-Informed Neural Networks}

\author{Maya Okawa}
\affiliation{
\institution{NTT Human Informatics Labs}
\city{Yokosuka}
\country{Japan}}
\email{maya.ookawa.af@hco.ntt.co.jp}

\author{Tomoharu Iwata}
\affiliation{
  \institution{NTT Communication Science Labs}
  \city{Kyoto}
  \country{Japan}
}
\email{tomoharu.iwata.gy@hco.ntt.co.jp}
\renewcommand{\shortauthors}{Maya Okawa \& Tomoharu Iwata}

\begin{abstract}
Opinion formation and propagation are crucial phenomena in social networks and have been extensively studied across several disciplines. Traditionally, theoretical models of opinion dynamics have been proposed to describe the interactions between individuals (i.e., social interaction) and their impact on the evolution of collective opinions. Although these models can incorporate sociological and psychological knowledge on the mechanisms of social interaction, they demand extensive calibration with real data to make reliable predictions, requiring much time and effort. Recently, the widespread use of social media platforms provides new paradigms to learn deep learning models from a large volume of social media data. However, these methods ignore any scientific knowledge about the mechanism of social interaction. In this work, we present the first hybrid method called Sociologically-Informed Neural Network (SINN), which integrates theoretical models and social media data by transporting the concepts of physics-informed neural networks (PINNs) from natural science (i.e., physics) into social science (i.e., sociology and social psychology). In particular, we recast theoretical models as ordinary differential equations (ODEs). Then we train a neural network that simultaneously approximates the data and conforms to the ODEs that represent the social scientific knowledge. In addition, we extend PINNs by integrating matrix factorization and a language model to incorporate rich side information (e.g., user profiles) and structural knowledge (e.g., cluster structure of the social interaction network). Moreover, we develop an end-to-end training procedure for SINN, which involves Gumbel-Softmax approximation to include stochastic mechanisms of social interaction. Extensive experiments on real-world and synthetic datasets show SINN outperforms six baseline methods in predicting opinion dynamics. 
\end{abstract}

\ccsdesc[300]{Applied computing~Sociology}
\ccsdesc[300]{Applied computing~Psychology}
\ccsdesc[500]{Computing methodologies~Neural networks}
\ccsdesc[500]{Information systems~Social networking sites}
\ccsdesc[500]{Mathematics of computing~Ordinary differential equations}
\ccsdesc[300]{Mathematics of computing~Stochastic differential equations}
\ccsdesc[100]{Mathematics of computing~Time series analysis}

\keywords{opinion dynamics; sequence prediction; social network}

\maketitle

\section{Introduction}

Opinion dynamics is the study of information and evolution of opinions in human society. 
In society, people exchange opinions on various subjects including political issues, new products, and social events (e.g., sports events). 
As a result of such interactions (i.e., social interaction), an individual's opinion is likely to change over time. 
Understanding the mechanisms of social interaction and predicting the evolution of opinions are essential 
in a broad range of applications such as business marketing \cite{DBLP:journals/access/Sanchez-NunezCH20,kumar2018detecting}, 
public voice control \cite{DBLP:conf/nldb/LaiPRR18}, friend recommendation on social networking sites \cite{eirinaki2013trust} and social studies \cite{moussaid2013social}. 
For example, predicting the evolution of opinion dynamics can help companies design optimal marketing strategies 
that can guide the market to react more positively to their products.  

Theoretical models called opinion dynamics models have been proposed for modeling opinion dynamics.  
The earliest work can be traced back to sociological and social psychological research by French \cite{french1956formal} and DeGroot \cite{degroot1974reaching}.  
The most representative ones are bounded confidence models \cite{DBLP:journals/jasss/HegselmannK02,DBLP:journals/advcs/DeffuantNAW00}, 
which take into account the psychological phenomenon known as confirmation bias --- 
the tendency of people to accept only information that confirms prior beliefs. 
These models specify the underlying mechanism of social interaction in the form of difference equations.  
Typically, opinion dynamics models are explored and validated by computer simulations such as agent-based models. 
The theoretical models are highly interpretable; and can utilize the wealth of knowledge from social sciences, especially in social psychology and sociology.
However, to learn predictive models, the computer simulations require calibration with real data, which incurs high manual effort and high computational complexity due to the massive data needed. 

Nowadays, social media is increasingly popular as means of expressing opinions \cite{DBLP:conf/icwsm/OConnorBRS10,devi2020literature} and they 
have become a valuable source of information for analyzing the formation of opinions in social networks.  
Data-driven methods have been applied to exploit large-scale data from social media for predicting the evolution of users' opinions. 
For instance, De \textsl{et al.} \cite{DBLP:conf/cikm/DeBBGC14} developed a simple linear model for modeling the temporal evolution of opinion dynamics.  
Some studies \cite{DBLP:conf/nips/DeVGBG16,DBLP:conf/icdm/KulkarniADBG17} proposed probabilistic generative models based on point processes 
that capture the non-linear dynamics of opinion formation. 
However, these models lack flexibility as they use hand-crafted functions to model social interaction.  
Recent work \cite{zhu2020neural} developed a deep learning approach that automatically learns 
the social interaction and underlying evolution of individuals' opinions from social media posts. 
While this approach enables flexible modeling of social interaction, 
it is largely agnostic to prior scientific knowledge of the social interaction mechanism. 

In this paper, we propose a radically different approach that integrates both large-scale data and prior scientific knowledge. 
Inspired by the recent success of physics-informed neural networks (PINNs) \cite{raissi2019deep}, 
we extend them to encompass opinion dynamics modeling. 
To this end, we first reformulate opinion dynamics models into ordinary differential equations (ODEs). 
We then approximate the evolution of individuals' opinions by a neural network. 
During the training process, by penalizing the loss function with the residual of ODEs that represent the theoretical models of opinion dynamics,  
the neural network approximation is made to consider sociological and social psychological knowledge. 

While PINNs have successfully incorporated the laws of physics into deep learning, 
some challenges arise when translating this method to opinion dynamics. 
First, most of them fail to capture the \textsl{stochastic nature} of social interaction. 
Many recent opinion dynamics models involve stochastic processes that emulate real-world interactions. 
For example, several bounded confidence models encode probabilistic relationships between individuals, in which individuals randomly communicate their opinion to others, via stochastic variables. 
Such stochastic processes require sampling from discrete distributions, which makes backpropagating the error through the neural network impossible.
Second, they cannot utilize \textsl{additional side information} including individuals' profiles, social connections (e.g., follow/following relationship in Twitter), and external influences from mass media, government policy, etc. 
In real cases, we can obtain rich side information from social media, including 
user profiles and social media posts (e.g., user profile description and tweet text attributes), 
which is described in natural language.  
Finally, they cannot consider \textsl{structural knowledge} on social interaction. 
Social interaction is represented by network structure, and each network has a certain cluster structure. 
Previous studies have empirically shown real-world social networks
consist of different user communities sharing same beliefs \cite{DBLP:conf/kdd/KumarNT06}. 

To address these challenges, we develop a new framework called \textsf{SINN} (Sociologically-Informed Neural Network), which is built upon PINNs. 
To take account of the \textsl{stochastic nature} of opinion dynamics models, 
we incorporate the idea of reparameterization into the framework of PINNs.  
Specifically, we replace the discrete sampling process with a differentiable function of parameters 
and a stochastic variable with fixed Gumbel-Softmax distribution. 
This makes the gradients of the loss propagate backward through the sampling operation and thus the whole framework can be trained end-to-end. 
Also, PINNs are broadened to include \textsl{rich side information}. 
To use users' profile descriptions, we combine the framework of PINNs and natural language processing techniques by adding a pre-trained language model.  
Moreover, building upon \textsl{structural knowledge} on the social interaction network (i.e., cluster structure), 
we apply low-rank matrix factorization to the parameters of ODEs. 
The proposal, \textsf{SINN}, incorporates the underlying sociology and social psychology, as described by ODEs, into the neural network 
as well as rich side information and structural knowledge, 
while at the same time exploiting the large amount of data available from social media.  

To evaluate our proposed framework we conduct extensive experiments on three synthetic datasets and three real-world datasets from social networks (i.e., Twitter and Reddit).  
They demonstrate that \textsf{SINN} provides better performance for opinion prediction than several existing methods, 
including classical theoretical models as well as deep learning-based models. 
To the best of our knowledge, this work is the first attempt to transpose the concepts of PINNs from natural science (i.e., physics) into social science (i.e., sociology and social psychology).
To foster future work in this area, % of multidisciplinary research, 
we will release our code and sample data on \path{https://github.com/mayaokawa/opinion_dynamics}. 

Our contributions are summarized as follows: 
\begin{itemize}
\item We propose the first hybrid framework called \textsf{SINN} (Sociologically-Informed Neural Network) that integrates a large amount of data from social media and underlying social science to predict opinion dynamics in a social network. 
\item \textsf{SINN} formulates theoretical models of opinion dynamics as ordinary differential equations (ODEs), 
and then incorporates them as constraints on the neural network. It also includes matrix factorization and a language model to incorporate rich side information and structural knowledge on social interaction.  
\item We propose an end-to-end training procedure for the proposed framework. To include stochastic opinion dynamics models in our framework, we introduce the reparameterization trick into \textsf{SINN}, which makes its sampling process differentiable.  
\item We conduct extensive experiments on three synthetic datasets and three real-world datasets from social networks. 
The results show that \textsf{SINN} outperforms six existing methods in predicting the dynamics of individuals' opinions in social networks. 
\end{itemize}

\section{Related Work}

\textbf{Physics-informed neural networks. }
From a methodological perspective, this paper is related to the emerging paradigm of physics-informed neural networks (PINNs) 
\cite{raissi2019deep}. 
PINNs incorporate physical knowledge into deep learning 
by enforcing physical constraints on the loss function, which is usually described by systems of partial differential equations (PDEs). These methods have been successfully applied to many physical systems, including  
fluid mechanics \cite{sun2020surrogate}, 
chemical kinetic \cite{ranade2019extended}, 
optics \cite{chen2020physics}. This paper is the first attempt to import this approach into the field of social science, especially for modeling opinion dynamics.  
This poses three major challenges. 
First, many common opinion dynamics models involve randomized processes; but most PINNs deal only with deterministic systems.
Very few works \cite{yang2020physics,zhang2019quantifying} explore stochastic PDEs with additive noise terms. 
However, they still fail to model stochastic terms that contain the model parameters. 
Second, they do not include rich side information such as profiles of social media users and social media posts.  
Third, they cannot take into account extra structural knowledge on social interaction.

\textbf{Opinion dynamics modeling. }
Opinion dynamics is the study of how opinions emerge and evolve through the exchange of opinions with others (i.e., social interaction). 
It has been a flourishing topic in various disciplines from social psychology and sociology to statistical physics and computer science. Theoretical models of opinion dynamics have been developed in divergent model settings with different opinion expression formats and interaction mechanisms. 
The first mathematical models can be traced back to works by sociologists and psychologists. 
In the 60s, French and John \cite{french1956formal} proposed the first simple averaging model of opinion dynamics,  
followed by Harary \cite{acemouglu2013opinion} and DeGroot \cite{degroot1974reaching}. 
Extensions to this line of work, such as Friedkin and Johnsen (FJ) model \cite{friedkin1990social}, 
incorporated the tendency of people to cling to their initial opinions. 
Bounded confidence models, including Hegselmann-Krause (HK) model \cite{DBLP:journals/jasss/HegselmannK02} 
and Deffuant-Weisbuch (DW) model \cite{DBLP:journals/advcs/DeffuantNAW00}, 
consider confirmation bias: the tendency of individuals to interact those with similar opinions. 
Later studies generalized bounded confidence models to account for multi-dimensional opinion spaces \cite{lanchier2012axelrod},  
asymmetric \cite{zhang2013opinion} and time-varying bounded confidence \cite{zhang2017opinion}. 
More recent works \cite{schweighofer2020agent,leshem2018impact,liu2013social} take into account stochastic interactions among individuals. 
Another line of work has adopted the concepts and tools of statistical physics to describe the behavior of interacting individuals, 
for example, the voter model \cite{DBLP:conf/ita/YildizPOS10}, the Sznajd model \cite{sznajd2000opinion}, and
the majority rule model \cite{galam1999application}. These physical models are built upon empirical investigations in social psychology and sociology.  
Opinion dynamics models have been tested by analytical tools as well as simulation models.  
Some works \cite{golub2012homophily,como2009scaling} employ differential equations for theoretical analysis. 
However, no analytic solutions of the differential equation are possible for most opinion dynamics models. 
Many other studies \cite{martins2008continuous,jager2005uniformity} adopt agent-based models to explore how the local interactions of individuals cause the emergence of collective public opinion in a simulated population of agents. 
However, to make reliable predictions, such simulations demand calibration with real data as done in \cite{sobkowicz2016quantitative,sun2013framework},  but the cost incurred in terms of time consumed and labor expended are high.  

Given the advances in machine learning, efforts have been made to utilize rich data collected from social media to model opinion dynamics. 
For instance, early work proposed a linear regression model, AsLM (asynchronous linear model) \cite{DBLP:conf/cikm/DeBBGC14}; 
it learns the linear temporal evolution of opinion formation. 
This model is based on a simple linear assumption regarding social interaction. 
SLANT \cite{DBLP:conf/nips/DeVGBG16} and its subsequent work, SLANT+ \cite{DBLP:conf/icdm/KulkarniADBG17}, 
apply a generative model based on multi-dimensional point processes to this task to capture the non-linear interaction among individuals. 
Unfortunately, these models still lack flexibility as deriving the point process likelihood demands excessively simple assumptions. 
Monti \textsl{et al.} \cite{DBLP:conf/kdd/MontiMB20} translate a classical bounded confidence model into a generative model and 
develop an inference algorithm for fitting that model to social media data. 
However, this model still relies on which opinion dynamics model is adopted. 
Moreover, it is intended to infer the individuals' opinions 
from the aggregated number of posts and discussions among social media users within a fixed time interval.  
Different from this model, this paper focuses on modeling the sequence of labeled opinions observed at different time intervals 
directly, without aggregation or inference. 
Recent work \cite{zhu2020neural} employs a deep learning model to estimate the temporal evolution of users' opinions from social media posts.  
Although this method can learn the influence of the other users' opinions through an attention mechanism, 
it is mostly data-driven and largely ignores prior knowledge about social interaction.

\section{Preliminaries}
We introduce the concept of opinion dynamics models, 
and then define the problem of predicting opinion dynamics.

\subsection{Opinion dynamics models}
Opinion dynamics models describe the process of opinion formation in human society. 
Conventional opinion dynamics models are equation-based, wherein one specifies the underlying mechanism of social interaction 
in the form of difference equations. Starting from DeGroot model \cite{degroot1974reaching},
many opinion dynamics models based upon theories in sociology and social psychology have been proposed with a wide variety of interaction mechanisms.  
In the following, we introduce some representative models of opinion dynamics.  

{\bf DeGroot model. } 
DeGroot model \cite{degroot1974reaching} is the most simple and basic model of opinion dynamics.  
In this model, each user $u$ holds opinion $x_u(t)\in[-1,1]$ on a specific subject (e.g., politics, product and event) at every time step $t$, 
where the extremal values correspond to the strongest negative (-1) and positive (1) opinions. 
The DeGroot model assumes that their opinions evolve in discrete time steps according to the following rule:
\begin{align}\label{eq:degroot}
x_u(t+1) = x_u(t) + \sum_{v\in\mathcal{U}\backslash u} a_{uv} x_v(t), 
\end{align}
where $\mathcal{U}$ is the set of users, 
$x_v(t)$ represents user $v$'s previous opinion, and 
$a_{uv}$ is the strength of the interactions between users $u$ and $v$.
In the DeGroot model, each user $u$ forms her opinion $x_u(t+1)$ at time step $t+1$ as the weighted sum of all previous opinions
at the preceding time step $t$. 
The DeGroot model captures the concepts of \textsl{assimilation} --- the tendency of individuals to move their opinions towards others.  
The continuous-time version of the DeGroot model was proposed by \cite{abelson1964mathematical}. 

{\bf Friedkin-Johnsen (FJ) model. } Friedkin and Johnsen \cite{friedkin1990social} extended the Degroot model to allow the users to have different susceptibilities to persuasion (i.e., the tendency to defer to others' opinions).  
According to the FJ model, the opinion at current time step $t$ is updated as the sum of her initial opinion $x_u(0)$ and all the previous opinions, 
which is denoted by: 
\begin{align}\label{eq:friedkin}
x_u(t+1) = s_u \sum_{v\in\mathcal{U}\backslash u} x_v(t) + (1-s_u) x_u(0), 
\end{align}
where the scaler parameter $s_u\in[0,1]$ measures user $u$'s susceptibility to persuasion.  
If $s_n=0$, user $u$ is maximally stubborn (attached to their initial opinion); on the other hand, $s_1=1$ indicates user $u$ is maximally open-minded. 
A continuous-time counterpart was proposed in \cite{taylor1968towards}. 

{\bf Bounded confidence model. } 
The bounded-confidence opinion dynamics model considers a well-known social psychological phenomenon: \textsl{confirmation bias} 
--- the tendency to focus on information that confirms one's preconceptions.
The most popular variant is the Hegselmann-Krause (HK) model.
The HK model assumes that individuals tend to interact only if their opinions lie within a bounded confidence interval $\delta$ of each other.  
Opinion update with bounded confidence interval $\delta$ is modeled as: 
\begin{align}\label{eq:hegselmann}
 \begin{aligned}
    x_u(t+1) &= x_u(t) + \vert N_u(t) \vert^{-1} \sum_{v\in N_u(t)} \big(x_v(t)-x_u(t)\big),
 \end{aligned}
\end{align}
with $N_u(t)=\{v\in\mathcal{U} : \, \vert x_u(t)-x_v(t)\vert \leq \delta\}$ being the set of users 
whose opinions fall within the bounded confidence interval of user $u$ at time step $t$,  
where $\vert N_u(x)\vert$ is the number of users with opinions within distance $\delta$.
Depending on the confidence parameter $\delta$ and the initial opinions, the interactions lead to different public opinion formations, 
namely, consensus (agreement of all individuals), polarization (disagreement between two competing opinions), 
or clustering (fragmentation into multiple opinions).  

\textbf{Stochastic bounded confidence model (SBCM). }
Recent works \cite{schweighofer2020agent,liu2013social,baumann2021emergence} extend the bounded confidence model by incorporating stochastic interactions based on opinion differences. These models define a distribution for the probability of interaction between users as a function of the magnitude of the separation between their opinions, 
and sample interaction $z_{uv}^t$ from it.  
The common form \cite{liu2013social,baumann2021emergence} is given by  
\begin{align}\label{eq:p_uv}
p(z_{uv}^t=1) = \frac{ \vert x_u(t)-x_v(t) \vert^{-\rho} }{ \sum_{v'} \vert x_u(t)-x_{v'}(t) \vert^{-\rho} },
\end{align}
where $p(z_{uv}^t=1)$ the probability of user $u$ selecting user $v$ as an interaction partner at time step $t$, and  
$\rho$ is the exponent parameter that determines the influence of opinion difference on the probabilities; 
$\rho>0$ means users with similar opinions are more likely to interact and influence each other, and $\rho<0$ means the opposite.

\subsection{Problem Definition}
We consider a social network with a set of users $\mathcal{U}=\{1,\ldots,U\}$, each of whom has an opinion on a specific subject, 
which evolves through social interaction.
We are given a collection of social media posts (e.g., tweets) that mention a particular object.  
Each post includes labeled opinion, either manually labeled or automatically estimated.  
Let there be $C$ labels, $\{1,...,C\}$, one corresponding to each opinion class.  
An opinion label can be, for example, an opinion category $y\in\{1,2,3\}$ 
indicating a ``positive'', ``neutral'' and ``negative'' opinion respectively.  
Formally, a social media post is represented as the triple $(u,t,y)$,
which means user $u$ made a post with opinion $y$, on a given subject matter, at time $t$. 
We denote $\mathcal{H}= \{(u_i,t_i,y_i)\}_{i=1}^I$ as the sequence of all posts made by the user up to time $T$. 

In addition to the sequence of opinions, we can have \textsl{additional side information}.  
For example, most social media services (e.g., Twitter) offer basic user information, 
such as the number of friends (followers) and profile description (i.e., a short bio). 
Here we consider the case of user profile descriptions.
Formally, such descriptions are represented by a sentence with $\vert d\vert$ words $d=\{w_1,...,w_{|d|}\}$. 
We denote a collection of user profiles by $\mathcal{D}=\{d^1,...,d^U\}$. 

Given the sequence of opinions $\mathcal{H}$ during a given time-window $[0, T)$ and user profiles $\mathcal{D}$, 
we aim to predict users' opinions   
at an arbitrary time $t^{\ast}$ in the future time window $[T, T+\Delta T]$.

\begin{figure*}[th]
  \includegraphics[width=0.73\linewidth]{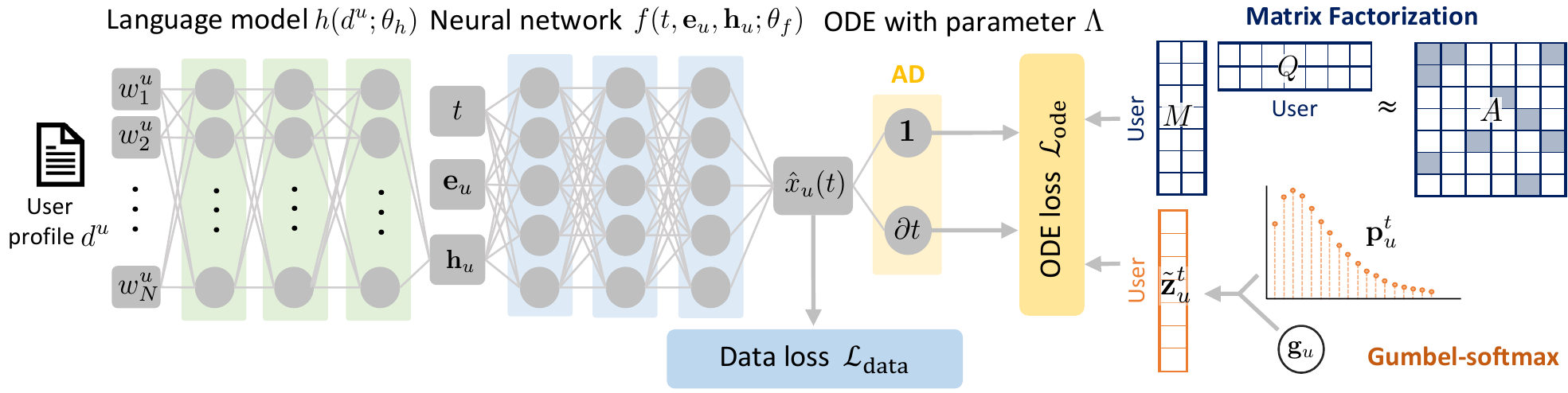}
\vspace{-2mm}
\caption{Overall architecture of our proposed method, \textsf{SINN} (Sociologically-Informed Neural Network). }\label{fig:overview} 
\vspace{-3.5mm}
\end{figure*}
\section{Proposed method}
In this work, we propose a deep learning framework, termed \textsf{SINN} (Sociologically-Informed Neural Network),  
for modeling and predicting the evolution of individuals' opinions in a social network. 

In this paper, inspired by recently developed physics-informed neural networks \cite{raissi2019physics},
we develop a deep learning framework that integrates both large-scale data from social media and prior scientific knowledge. 
\cref{fig:overview} shows the overall architecture of \textsf{SINN}. 
Opinion dynamics models, in their most common form, are expressed by difference equations (e.g., Equations~\eqref{eq:degroot}~to~\eqref{eq:p_uv}). 
We first reformulate these difference equations of opinion dynamics as ordinary differential equations (ODEs), 
which are shown in \textcolor[RGB]{246,195,28}{yellow} in \cref{fig:overview}.  
We then approximate the evolution of individuals' opinions by a neural network (denoted by \textcolor[RGB]{136,193,219}{light blue}). 
By penalizing the loss function with the ODEs, the neural network is informed by a system of ODEs 
that represent the opinion dynamics models; a concept drawn from theories of sociology and social psychology.  
The proposed framework is further combined with a language model (colored \textcolor[RGB]{133,160,102}{green} in \cref{fig:overview}) 
to fuse rich side information present in natural language texts (e.g., user profile descriptions).  
We also utilize low-rank approximation (\textcolor[RGB]{0,0,139}{dark blue}) by factorizing the ODE parameters to capture the cluster structure of the social interaction network.
The parameters of our model are trained via backpropagation.  

Several opinion dynamics models involve stochastic processes,  
which encode random interactions between individuals. 
Stochastic sampling poses a challenge for end-to-end training because it would prevent the gradient of the loss from being back-propagated.  
In order to back-propagate through the stochastic variables, we leverage the reparametrization trick (\textcolor[RGB]{216,137,0}{orange} in \cref{fig:overview}) 
commonly used for gradient estimation \cite{kusner2016gans}.  
This formulation makes the sampling process differentiable and hence the whole scheme can be trained in an end-to-end manner. 

We elaborate the formulation of \textsf{SINN} in \cref{sec:formulation}, 
followed by parameter learning in \cref{sec:learning}. 
The prediction and optimization procedure is presented in \cref{sec:app_proposed}.

\subsection{Model Formulation}\label{sec:formulation}

\subsubsection{ODE constraint. }
To incorporate social scientific knowledge, we rewrite opinion dynamics models 
as ordinary differential equations (ODEs). 
In the following, we formulate the ODEs by using some representative opinion dynamics models as guiding examples. 
Note though that our proposed framework can be generalized to any other opinion dynamics model. 
We use $\Lambda$ to denote the parameters of ODE. 

\textbf{DeGroot model. } For the DeGroot model \cite{degroot1974reaching}, 
we can transform the difference equation of \cref{eq:degroot} into an ODE as follows: 
\begin{align}\label{eq:ode_degroot0}
\frac{dx_u(t)}{dt} = \sum_{v\in\mathcal{U}\backslash u} a_{uv} x_v(t), 
\end{align}
where $t$ is time and $x_u(t)$ is the user $u$'s latent opinion at time $t$.  
Higher values of $x_u(t)$ represent more positive opinions towards a given topic. 
Here we replace the discrete index $t$ (representing time step) of \cref{eq:degroot} by the continuous variable $t$ (representing continuous time).  
The weight parameter, $a_{uv}$, indicates how much user $u$ is influenced by the opinion of user $v$. 
\cref{eq:ode_degroot0} postulates that an individual's opinion evolves over $t$ as a result of being influenced by the opinions of others. 
We can further modify \cref{eq:ode_degroot0} to capture \textsl{structural knowledge} on social interaction (i.e., cluster structure). 
Real-world social networks have been found to contain multiple groups/communities of users sharing a common context \cite{DBLP:conf/kdd/KumarNT06}.  %sharing same beliefs and opinions
To discover the cluster structure of the social interaction network, 
we apply matrix factorization to the matrix of weight parameter $A=\{a_{uv}\}\in\mathbb{R}^{U\times U}$. 
In particular, we factorize matrix $A$ into the two latent matrices of $M\in\mathbb{R}^{U\times K}$ and $Q\in\mathbb{R}^{U\times K}$: $A \approx M^{\top}Q$,
where $K\ll U$ is the dimension of the latent space. 
Based on this parametrization, \cref{eq:ode_degroot0} reduces to  
\begin{align}\label{eq:ode_degroot}
\frac{dx_u(t)}{dt} = \sum_{v\in\mathcal{U}\backslash u} \textbf{m}_u^{\top} \textbf{q}_v x_v(t), 
\end{align}
where $\textbf{m}_u\in\mathbb{R}^{K}$ and $\textbf{q}_v\in\mathbb{R}^{K}$ correspond to 
the $u$-th column and $v$-th column of $M$ and $Q$. 
\cref{eq:ode_degroot} accounts for \textsl{assimilation}, a well-known phenomenon in social psychology.  

\textbf{Friedkin-Johnsen (FJ) model. } 
For the FJ model \cite{friedkin1990social}, 
the difference equation of \cref{eq:friedkin} can be transferred 
into an ODE of the form:  
\begin{align}\label{eq:ode_fj}
\frac{x_u(t)}{dt} = s_u \sum_{v\in\mathcal{U}\backslash u} x_v(t) + (1-s_u) x_u(0)- x_u(t), 
\end{align}
where $x_u(0)$ is the innate opinion of user $u$ at time $t=0$.  
In implementing this, we set the first expressed opinion of each user $u$ as the innate opinion $x_u(0)$. 
The interaction weight $a_{uv}$ denotes the influence user $u$ puts on $v$. 
The scaler parameter, $s_u\in[0,1]$, controls user $u$'s susceptibility to persuasion and so decides the degree of which the user is stubborn or open-minded. 

\textbf{Bounded confidence model (BCM). }
We transform the bounded confidence model of Hegselmann and Krause \cite{DBLP:journals/advcs/DeffuantNAW00} into an ODE as follows. 
Since the original model function (\cref{eq:hegselmann}) is not differentiable, we replace the threshold function in \cref{eq:hegselmann} with a sigmoid function.  
Formally, 
\begin{align}\label{eq:ode_bcm}
\frac{dx_u(t)}{dt} = \sum_{v\in\mathcal{U}} \sigma\big(\delta-|x_u(t)-x_v(t)|\big) \big(x_v(t)-x_u(t)\big), 
\end{align}
where $\sigma(z)=(1+e^{-\gamma z})^{-1}$ denotes a sigmoid function. 
As the slope of sigmoid function $\gamma$ approaches infinity, \cref{eq:ode_bcm} will approach the thresholding function in the original model (\cref{eq:hegselmann}).  
This model choice draws on the theory of \textsl{confirmation bias}: Users tend to accept opinions that agree with theirs while discarding others. 
In \cref{eq:ode_bcm}, if two users are close enough in their opinions (within confidence threshold $\delta$), 
they are likely to interact and develop a closer opinion;  
otherwise, they are not likely to interact, and maintain their original opinions.

\textbf{Stochastic bounded confidence model (SBCM). }
The stochastic bounded confidence model (SBCM) involves sampling $z_{uv}^t\sim p(z_{uv}^t)$ in \cref{eq:p_uv},  
which emulates stochastic interactions between users.  
Before triggering end-to-end training, we have to propagate the gradient through 
the non-differentiable sampling operation $\sim$ from a discrete distribution. Notice that as described in \cref{sec:nn}, latent opinion $x_u(t)$ is modeled by a neural network that involves a set of parameters to be optimized.  
To deal with the problem of the gradient flow block, we leverage the reparameterization approach.  
Using the Gumbel-Softmax trick \cite{jang2016categorical}, 
we can replace the discrete samples ${\bf z}_u^t=\{z_{u1}^t,\ldots,z_{uU}^t\}$ 
with differentiable one-hot approximation $\tilde{\bf z}_u^t=\{\tilde{z}_{u1}^t,\ldots,\tilde{z}_{uU}^t\}\in\mathbb{R}^U$ as follows:  
\begin{align}\label{eq:haty_u}
\tilde{\bf z}_{u}^t = \text{Softmax} \big( [ \log{({\bf p}_{u}^t)} + {\bf g}_{u} ] / \tau \big),  
\end{align}
where ${\bf p}_u^t\in\mathbb{R}^U$ consists of probability $p(z_{uv}^t)$ for a set of users $v\in\mathcal{U}$ defined in \cref{eq:p_uv}. 
${\bf g}_u\in\mathbb{R}^U$ is the noise vector whose element is i.i.d and generated from the Gumbel distribution\footnote{Gumbel(0,1) distribution can be sampled using inverse transform sampling \cite{jang2016categorical} by drawing $\mu \sim\text{Uniform}(0,1)$ and then computing $g = -\log(-\log(\mu))$}.  
This formulation allows randomness $\textbf{g}_{u}$ to be separated from the deterministic function of the parameters. 
In \cref{eq:haty_u}, we relaxed the non-differentiable argmax operation to the continuous softmax function.
$\tau$ is a temperature parameter controlling the degree of approximation;   
when $\tau\rightarrow 0$, the softmax function in \cref{eq:haty_u} approaches argmax and ${\bf z}_u^t$ becomes a one-hot vector. With the above formulation, 
we can propagate gradients to probabilities $p_{uv}^t$ through the sampling operation. 
Finally, SBCM is rewritten as 
\begin{align}\label{eq:ode_sbcm}
\frac{dx_u(t)}{dt} = \sum_{v\in\mathcal{U}} \tilde{z}_{uv}^t \big(x_v(t)-x_u(t)\big).  
\end{align}

\subsubsection{Neural network. }\label{sec:nn}
We then construct a neural network for approximating the evolution of individuals' opinions. 
In this paper, we employ a feedforward neural network (i.e., FNN) with $L$ layers. 
The neural network takes time $t$ and user $u$ as inputs and outputs latent opinion $x_u(t)$ of user $u$ at that time, which is denoted by,  
$x_u(t) \approx f(t, {\bf e}_u; \theta_f)$, 
where $f(\cdot)$ represents the neural network, ${\bf e}_u$ is the one-hot encoding of a user index $u$, 
and $\theta_f$ represents the trainable parameters, namely, a set of weights and biases. 
We denote the output of neural network as $\hat{x}_u(t)=f(t, \textbf{e}_u; \theta_f)$. 
In our implementation, we adopt {\sl tanh} activation for all layers. During training, we enforce the FNN output $\hat{x}_u(t)$ to (i) reproduce the observed opinions; and 
(ii) satisfy the governing ODEs that represent the theoretical models of opinion dynamics. 

The FNN can be extended to incorporate rich side information.  
Here we assume that we are given individuals' profile descriptions. 
To extract meaningful features from the profile descriptions, we adopt natural language processing techniques. 
We obtain hidden user representation $\textbf{h}_u=h(d^u,\theta_h)$, where $h(\cdot)$ is the language model, $d^u=\{w_1^u,...,w_{|d^u|}^u\}$ is the sequence of words, 
$\theta_h$ is the trainable parameters. 
In this work, we adopt a pre-trained BERT Transformer \cite{devlin2018bert} as the language model 
and fine-tune an attention layer built on top of the pre-trained BERT model. 
The architecture of the language model is detailed in \cref{sec:app_language}.  
The hidden user representation is embedded into the FNN: 
$x_u(t) \approx f(t, \textbf{e}_u, \textbf{h}_u; \theta_f)$. 
Although different types of additional information may be available, 
the above formulation focuses on user profile descriptions.  
Notice that our formulation is flexible enough to accept any new information.  
For example, the FNN can be further integrated with deep learning models %such as LSTM   
to capture the influence of mass media (e.g., news websites) from their contents (e.g., news articles).  

%%%%%%%%%%%%%%%%
{\small\begin{table*}[!t]
\caption{Statistics of the three real-world datasets used in this paper. }
\vspace{-2.5mm}
\begin{tabular}{lcccccc} \toprule
                 & \# Users & \# Posts & \# Positive & \# Negative & \# Classes & Time span \\ \midrule
Twitter BLM & 1,721 & 16,858 & 12,072 & 4,786 & 4 & August 01, 2020 - March 31, 2021 \\
Twitter Abortion & 2,359 & 24,217 & 8,822 & 15,395 & 4 & June 01, 2018 - April 30, 2021 \\
Reddit Politics & 1,335 & 22,563 & 5,724 & 16,839 & 2 & April 30, 2020 - November 03, 2020 \\
\bottomrule 
\end{tabular}\label{tab:stats}
\vspace{-2mm}
\end{table*}}
%%%%%%%%%%%%%%%%

\subsection{Parameter Learning}\label{sec:learning}
The proposed method, \textsf{SINN}, is trained to approximate the observations of opinions 
while satisfying the sociological and social psychological constraints determined by the ODEs. We aim to learn the neural network parameters $\theta_f$, 
the language model parameters $\theta_h$, as well as the ODE parameters $\Lambda$.  
The total loss $\mathcal{L}$ is thus composed of the data loss $\mathcal{L}_{\text{data}}$, 
the ODE loss $\mathcal{L}_{\text{ode}}$,  
and a regularization term, which is defined by
{\small\begin{align}\label{eq:totalloss}
\mathcal{L}(\theta_f, \theta_h, \Lambda; \mathcal{H}, \mathcal{D}) = 
   \mathcal{L}_{\text{data}}(\theta_f,\theta_h; \mathcal{H}, \mathcal{D})  
   + \alpha \mathcal{L}_{\text{ode}}(\theta_f,\theta_h, \Lambda) 
      + \beta \mathcal{R}(\Lambda),
\end{align}}
\hspace{-1mm}where $\mathcal{L}_{\text{data}}$ measures the discrepancy between the labeled opinion and the neural network output. 
Meanwhile, $\mathcal{L}_{\text{ode}}$ measures the residual of ODEs, enforcing given sociological and social psychological constraints described by the ODEs.
Hyperparameter $\alpha$ controls the balance between observed data $\mathcal{H}$ and prior scientific knowledge.  
$\mathcal{R}(\Lambda)$ is a regularization term and $\beta$ is the regularization parameter. 
In our experiment, the trade-off hyperparameters $\alpha$ and $\beta$ were chosen using grid search. 
We elaborate on each term of the loss function below.  

{\bf Data Loss. }
The data loss term measures the deviation of predicted opinion $\hat{y}_i$  
from ground truth opinion label $y_i$.  
We project latent opinion $\hat{x}_{u_i}(t_i)$ of user $u_i$ at time $t_i$ to a probability distribution over opinion class labels by: 
$\hat{y}_i = \sigma ( \hat{x}_{u_i}(t_i) \textbf{w}_{\sigma} + \textbf{b}_{\sigma} )$, 
where $\textbf{w}_{\sigma}\in\mathbb{R}^C$ is the scale parameter, 
$\textbf{b}_{\sigma}\in\mathbb{R}^C$ is the bias parameter, and  
$\sigma$ is a mapping function (softmax or sigmoid activation function for multi-class and two-class opinion labels, respectively).  
We adopt cross-entropy loss as the data loss 
to measure the difference between the predicted label distribution $\hat{y}_i$ with the truth label of opinion $y_i$: 
{\small\begin{align}\label{eq:data_loss}
\mathcal{L}_{\text{data}}(\theta_f,\theta_h; \mathcal{H}, \mathcal{D}) = \frac{1}{I}\sum_{i=1}^{I} L_{\text{CE}}\big( \hat{y}_i, y_i \big), 
\end{align}}
\hspace{-1mm}where $L_{\text{CE}}(\cdot)$ denotes cross-entropy loss. 
\cref{eq:data_loss} ensures that the neural network output coincides with the ground-truth opinion labels.

{\bf ODE Loss. }
This ODE loss enforces given scientific constraints as described by the ODEs (\cref{eq:ode_degroot,eq:ode_fj,eq:ode_bcm,eq:haty_u,eq:ode_sbcm}). 
To ensure the difference between the left side and right side of all ODEs are equal to zero, 
we minimize the mean squared error between them. 
Taking the DeGroot model in \cref{eq:ode_degroot} as example, ODE loss is written as  
{\small\begin{align}\label{eq:loss}
\mathcal{L}_{\text{ode}}(\theta_f,\theta_h, \Lambda) = \frac{1}{J}\sum_{j=1}^J \sum_{u\in\mathcal{U}} 
\left( \left. \frac{d\hat{x}_{u}(t)}{dt} \right|_{t=\tau_j} - \sum_{v\in\mathcal{U}\backslash u} \textbf{w}_{u}^{\top} \textbf{h}_v \hat{x}_v(\tau_j) \right)^2,
\end{align}}
\hspace{-1mm}where $\{\tau_1,...,\tau_J\}$ is a set of $J$ collocation points.
We randomly generate the collocation points from the time domain $\tau_j\in[0,T+\Delta T]$.  
$\mathcal{L}_{\text{ode}}$ measures the discrepancy of the ODEs at the $J$ collocation points,  
which ensures that the neural network output satisfies the ODE constraints.  
The first derivative term $\frac{d\hat{x}_{u}(t)}{dt}$ in \cref{eq:loss} can be easily computed using automatic differentiation \cite{baydin2018automatic} 
in deep learning packages such as PyTorch \cite{paszke2017automatic}. 

{\bf Regularizer. }
We can use auxiliary constraints as the regularization term $\mathcal{R}(\cdot)$ in \cref{eq:totalloss}. 
For DeGroot model, to prevent over-fitting, we add the following regularization on the latent matrices: 
$\mathcal{R}(\Lambda) =  \Vert M \Vert_1 + \Vert Q \Vert_1$, 
where $\Vert M \Vert_1$ denotes the $l_1$-norm of the latent matrix $M$. 

The parameter optimization is described in \cref{sec:app_opt}.

\section{Experiments}
We conduct qualitative and quantitative experiments to evaluate the performance of the proposed method, \textsf{SINN} (Sociology-Informed Neural Network), 
on both synthetic and real-world datasets. 
In this section, we present our experimental settings and results.  
Our source code and sample data are publicly available\footnote{\path{https://github.com/mayaokawa/opinion_dynamics}}. 

\subsection{Datasets}
Experiments were conducted on three synthetic datasets of different settings and three real-world datasets collected from Twitter and Reddit. 
The data preprocessing procedure is detailed in \cref{sec:datasets}.

\subsubsection{Synthetic datasets. } 
We generate synthetic datasets using the stochastic bounded confidence model (SBCM). 
We simulate the following three scenarios created with three different exponent parameters $\rho$ of the SBCM in \cref{eq:p_uv}: 
i) Consensus ($\rho=-1.0$), 
ii) Polarization ($\rho=0.5$), and iii) Clustering ($\rho=0.05$).  
The bottom panel of \cref{fig:network_synthetic} shows the simulated opinion evolution for the Consensus dataset.  
Our simulation involves $U=200$ users whose initial opinions are uniformly distributed. 
All simulations were run for $T=200$ timesteps. 
For a fair comparison, we converted the simulated continuous opinion values into five discrete labels and used these labels as the input of the models.  
Please refer to \cref{sec:synthetic_dataset} for details.  

\subsubsection{Real datasets. } 
We use two Twitter datasets and one Reddit dataset. 
\cref{tab:stats} shows the statistics of the three real-world datasets. 

\textbf{Twitter datasets. } 
We construct two \textsl{Twitter datasets} by collecting tweets related to specific topics, i.e., {\sl BLM} (BlackLivesMatter) and {\sl Abortion},  
through the Twitter public API\footnote{\label{note:twitter}https://developer.twitter.com/en/docs. Accessed on May 15, 2021.} using hashtags and keywords. 
For {\sl BLM}, we crawled tweets containing hashtag \#BlackLivesMatter from 
%August 1, 2020, to March 1, 2021. 
August 1, 2020, to March 31, 2021. 
For {\sl Abortion} dataset, tweets were extracted between 
June 1, 2018, and April 30, 2021, 
using hashtag \#Abortion. Each tweet contains date/time, user, and text.
To remove bots, and spam users, 
we filtered the set of users based on the usernames and the number of posts. The preprocessing procedure is presented in \cref{sec:twitter}. 
All tweets are manually annotated into highly positive, positive, negative, highly negative towards the corresponding topic. 
Along with the tweets, we obtained the \textsl{profile descriptions} (i.e., the user-defined texts describing their Twitter accounts) for all the users.  
For each user, the profile description is represented as a collection of words.  

\textbf{Reddit datasets. } 
For Reddit dataset, we collect posts from two Reddit communities (i.e., subreddits) for political discussion: 
\texttt{r/libertarian} and \texttt{r/conservative}. 
All submissions were retrieved through the Pushshift API\footnote{\label{note:reddit}https://pushshift.io/. Accessed on April 30, 2021. } from April through November 2020, 
and include date/time, user, and subreddit.
Users with less than five posts and more than 1,000 posts were discarded, leaving 1,335 users.
We label posts in \texttt{r/libertarian} 0 (negative towards conservatism) and \texttt{r/conservative} 1 (positive).

\subsection{Comparison methods}\label{sec:baselines}
We compare the prediction performance of \textsf{SINN} with the following baselines: 
\begin{itemize}
\item \textsf{Voter} (Voter model) \cite{DBLP:conf/ita/YildizPOS10}: \textsf{Voter} is one of the simplest models of opinion dynamics. 
At each time step, users select one of the users uniformly at random and copy her opinion. 
\item \textsf{DeGroot} (DeGroot model) \cite{degroot1974reaching}: \textsf{DeGroot} is a classical opinion dynamics model; it assumes users update their opinions iteratively to the weighted average of the others' opinions. 
We analytically solve \cref{eq:ode_degroot0} and fit the model parameter $a_{uv}$.  
\item \textsf{AsLM} (Asynchronous Linear Model) \cite{DBLP:conf/cikm/DeBBGC14}: \textsf{AsLM} 
is a linear regression model where each opinion is regressed on the previous opinions. 
\item \textsf{SLANT} \cite{DBLP:conf/nips/DeVGBG16}: \textsf{SLANT} is a temporal point process model 
that captures the influence of the other users' opinions. 
It characterizes a self-exciting mechanism of the interaction between users and its time decay is based on intensity.  
\item \textsf{SLANT+} \cite{DBLP:conf/icdm/KulkarniADBG17}: \textsf{SLANT+} is an extension of SLANT, which combines recurrent neural network (RNN) with a point process model.  
RNN learns the non-linear evolution of users' opinions over time.
\item \textsf{NN}: \textsf{NN} is the proposed method without ODE loss $\mathcal{L}_{\text{ode}}$ in \cref{eq:totalloss}, 
in which the trade-off hyperparameters are fixed to $\alpha=0$ and $\beta=0$. 
\end{itemize}

Detailed settings for the baselines are specified in \cref{sec:baseline_settings}. 

\subsection{Experimental Setup}

\subsubsection{Evaluation Protocol. }
For each synthetic dataset, we split the training, validation, and test set in a proportion of 50\%, 20\%, 30\% in chronological order.  
We divide each real-world dataset into train, validation, and test set with ratios of 70\%, 10\%, and 20\%. 
To evaluate the prediction performance of all models, we measure the accuracy (ACC) and the macro F1 score (F1),
both of which assess the agreement between predicted opinion classes and the ground truth.

\subsubsection{Hyperparameters. }
Hyperparameters of all methods, including comparison methods and ours, are tuned by grid search on the validation set.
For the neural network models, we determine the number of layers $L$ in the range of $\{3,5,7\}$
and the hidden layer dimension in the range of $\{8,12,16\}$.  
For our \textsf{SINN}, we search in $\{0.1,1.0,5.0\}$ for the trade-off hyperparameters $\alpha$ and $\beta$; 
$\{1,2,3\}$ is taken as the dimension of the latent space $K$. 
We test four opinion dynamics models (Equations~\eqref{eq:ode_degroot}~to~\eqref{eq:ode_sbcm}) and report the results of the best one. 
We selected the pre-trained $\text{BERT}_\text{BASE–uncased}$ as the language model of our \textsf{SINN} 
(See \cref{sec:impl_detail} for the detailed setting). 
We use the Adam optimizer \cite{kingma2014adam} with learning rate 0.001 for all experiments.

{\small\begin{table}[!t]
\caption{F1 score (F1) and Accuracy (ACC) 
for predicting opinions from three synthetic datasets. 
Higher scores indicate better prediction performance.  
The best performance is highlighted in bold. % Our \textsf{SINN} outperforms the six baselines. 
}
\vspace{-2.5mm}
\begin{tabular}{lcccccc} \toprule
                    & \multicolumn{2}{c}{Consensus} & \multicolumn{2}{c}{Polarization} & \multicolumn{2}{c}{Clustering} \\ 
\cmidrule(lr){2-3} \cmidrule(lr){4-5}  \cmidrule(lr){6-7}
                    & ACC & F1 & ACC & F1 & ACC & F1  \\ \midrule
\textsf{ Voter    } & 0.457 & 0.197 & 0.248 & 0.204 & 0.274 & 0.202 \\
\textsf{ DeGroot  } & 0.458 & 0.260 & 0.706 & 0.515 & 0.765 & 0.524 \\
\textsf{ AsLM     } & 0.155 & 0.264 & 0.113 & 0.079 & 0.477 & 0.325 \\
\textsf{ SLANT    } & 0.006 & 0.002 & 0.038 & 0.025 & 0.190 & 0.064 \\
\textsf{ SLANT+   } & 0.008 & 0.003 & 0.113 & 0.041 & 0.386 & 0.111 \\
\textsf{ NN       } & 0.771 & 0.426 & 0.603 & 0.451 & 0.875 & 0.817 \\
\textsf{ Proposed } & \textbf{ 0.794} & \textbf{ 0.574} & \textbf{ 0.761} & \textbf{ 0.738} & \textbf{ 0.895} & \textbf{ 0.843} \\
\bottomrule 
\end{tabular}\label{tab:error_syn}
\vspace{-2mm}
\end{table}}

{\small\begin{table}[!t]
\caption{F1 score (F1) and Accuracy (ACC) 
for predicting opinions from three real-world datasets.  
Higher is better. 
The best performance is highlighted in bold. % Our \textsf{SINN} outperforms the six baselines. 
}
\vspace{-2.5mm}
\begin{tabular}{lcccccc} \toprule
                    & \multicolumn{2}{c}{Twitter BLM} & \multicolumn{2}{c}{Twitter Abortion} & \multicolumn{2}{c}{Reddit Politics} \\ 
\cmidrule(lr){2-3} \cmidrule(lr){4-5}  \cmidrule(lr){6-7}
                    & ACC & F1 & ACC & F1 & ACC & F1  \\ \midrule
\textsf{ Voter    } & 0.199 & 0.163 & 0.222 & 0.170 & 0.628 & 0.500 \\
\textsf{ DeGroot  } & 0.203 & 0.131 & 0.358 & 0.203 & 0.807 & 0.389 \\
\textsf{ AsLM     } & 0.092 & 0.117 & 0.435 & 0.195 & 0.789 & 0.441 \\
\textsf{ SLANT    } & 0.105 & 0.070 & 0.425 & 0.175 & 0.733 & 0.496 \\
\textsf{ SLANT+   } & 0.091 & 0.042 & 0.437 & 0.152 & 0.789 & 0.441 \\
\textsf{ NN       } & 0.336 & 0.237 & 0.441 & 0.369 & 0.875 & 0.824 \\
\textsf{ Proposed } & \textbf{ 0.359} & \textbf{ 0.246} & \textbf{ 0.467} & \textbf{ 0.412} & \textbf{ 0.927} & \textbf{ 0.884} \\
\bottomrule 
\end{tabular}\label{tab:error_real}
\vspace{-2mm}
\end{table}}

\subsection{Quantitative Evaluation}
We first report the prediction performance of different methods in terms of the two evaluation metrics on the three synthetic datasets in \cref{tab:error_syn}. 
We can observe that the proposed \textsf{SINN} achieves the best results in terms of accuracy (ACC) and F1 score (F1). 
None of the comparison methods perform robustly across all datasets. 
\textsf{AsLM} achieves the second best F1 score among the baselines for the Consensus dataset 
but fails on the Polarization and Clustering datasets. 
\textsf{DeGroot} outperforms the other baseline methods on the Polarization dataset, 
while providing relatively poor performance for the Consensus and Clustering datasets.  
It is due to the fact that they depend on the specific choice of opinion dynamics model used.  
In contrast, \textsf{SINN} is general and suits different opinion dynamics models, which yields better performance in all experiments. %all four 
In an additional experiment (\cref{sec:choice_odm}), we demonstrated that \textsf{SINN} can be used to select one of a set of opinion dynamics models 
that best captures hidden social interaction underlying data.  

\cref{tab:error_real} shows the results of the existing methods and \textsf{SINN} on three real-world datasets.  
\textsf{SINN} achieves the best prediction performance in all cases.
\textsf{AsLM} gives almost the worst performance for Twitter BLM dataset. 
This is because \textsf{AsLM} is a linear model and cannot adequately depict the complex interaction between users.  
\textsf{SLANT} and \textsf{SLANT+} have less predictive power since 
they assume a fixed parametric form for social interaction.  
Compared with these methods, deep learning-based models (i.e., \textsf{NN} and \textsf{SINN}) offer improved performance in terms of accuracy and F1 score. 
The reason is that the deep learning-based models can learn the complex mechanisms of opinion formation in real-world social networks 
due to the influence of external factors (e.g., mass media) and random behavior,  
which are not considered in the other methods.  
Importantly, \textsf{SINN} yields better prediction performance than \textsf{NN} across all the datasets.  
Compared with \textsf{NN}, \textsf{SINN} achieves 11.7\% improvement in terms of F1 for Twitter Abortion dataset. 
For Reddit dataset, it produces a result of 88.4\% in F1, which is an 7.3\% improvement over \textsf{NN}. 
This result verifies the effectiveness of incorporating prior scientific knowledge into the deep learning model, 
as well as the advantage of our model design.  
From these results, we can conclude that \textsf{SINN} can learn from both data and theoretical models. 
We perform a parameter study in \cref{sec:sensitivity}.

\begin{figure}[!t]
  \includegraphics[width=0.9\linewidth]{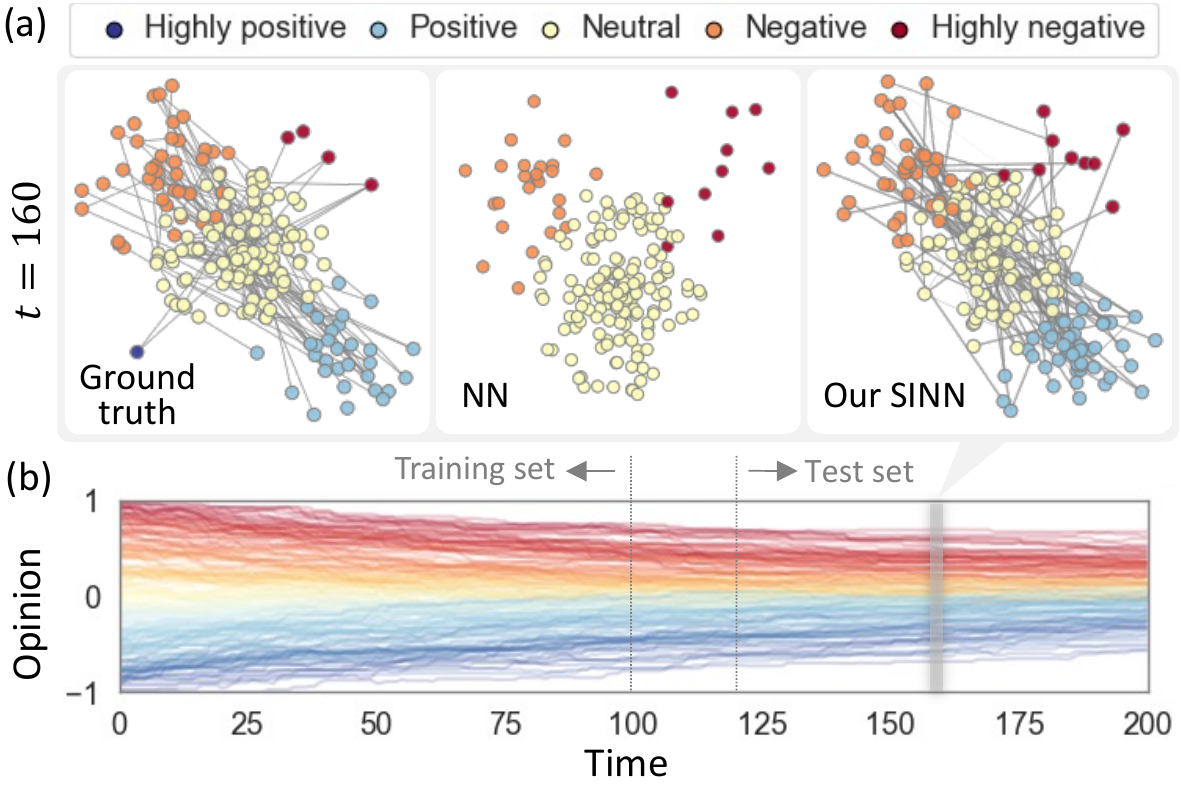}
\vspace{-4mm}
\caption{
(a) Interactions between users at time $t=160$, 
which is simulated (left) and predicted by \textsf{NN} (middle) and \textsf{SINN} (right). 
Nodes represent users where the color indicates their opinion classes.  
The left graph illustrates the true social interactions during $t\in[160,170]$. 
The middle graph shows the opinion classes predicted by \textsf{NN}. 
The right graph visualizes the opinion classes and the interaction parameter $\tilde{z}_{uv}^t$ of SBCM (\cref{eq:ode_sbcm}) learned by \textsf{SINN}. 
(b) Temporal evolution of individuals' opinions simulated in Consensus dataset. 
Each curve represents one individual. 
}\label{fig:network_synthetic} 
\vspace{-5mm}
\end{figure}

\subsection{Qualitative Evaluation via Case Studies}
% To illustrate the performance of our model intuitively, we compared the real values with the predicted values, shown in Figures 7 and 8, respectively.
\cref{fig:network_synthetic} depicts the temporal changes in population opinions and the networks of social interactions for the Consensus dataset. 
In the Consensus dataset, the population reaches consensus (\cref{fig:network_synthetic}(b)) 
through social interactions between users with opposite opinions (\cref{fig:network_synthetic}(a) left). 
We can see in \cref{fig:network_synthetic}(a) that \textsf{SINN} (right) can duplicate the actual interactions 
that frequently occur between opposing users (left). 
As can be seen from the node colors, \textsf{SINN} (right) better captures the actual opinion dynamics (left) than \textsf{NN} (middle). 
In \textsf{NN}, a large number of negative samples (light blue nodes) are mislabeled as neutral ones (yellow). 
\textsf{SINN}, however, misclassifies only a few negative samples as neutral. This result validates the benefit of incorporating prior sociological and social psychological knowledge in the form of the ODEs.  

\begin{figure}[!t]
  \includegraphics[width=0.89\linewidth]{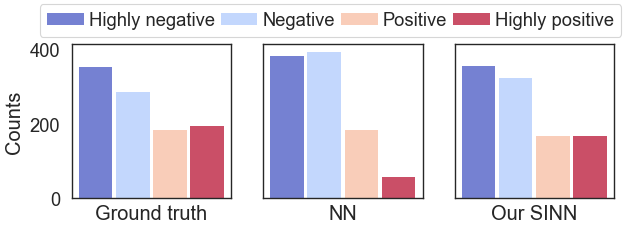}
\vspace{-3mm}
\caption{Distribution of opinion classes on November 7, 2020 for Twitter Abortion dataset. 
}\label{fig:hist_real} 
\vspace{-5mm}
\end{figure}
\cref{fig:hist_real} compares the distribution of opinion classes for Twitter Abortion dataset. 
For comparison, we show actual and predicted class distributions on November 7, 2020. 
We can see that \textsf{SINN} better reproduces the ground-truth label distribution than \textsf{NN}.  
On the other hand, \textsf{NN} underestimates the number of highly negative samples (red bar).  
This further emphasizes the importance of using prior knowledge and the effectiveness of our approach. 

\begin{figure}[!t]
  \subfigure[\textsf{NN}]{\centering\includegraphics[width=0.43\linewidth]{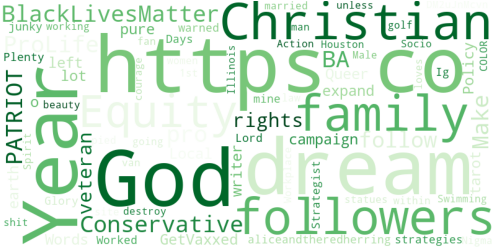}\label{fig:attention_NN}}\hspace{4mm}
  \subfigure[Our \textsf{SINN}]{\centering\includegraphics[width=0.43\linewidth]{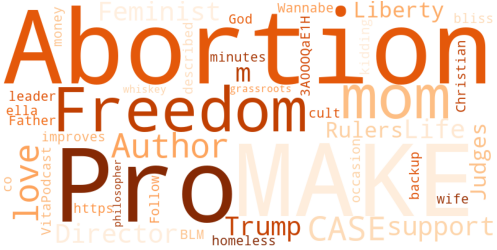}\label{fig:attention_Proposed}}
\vspace{-4mm}
\caption{ 
Important words in profile descriptions identified by attention mechanisms in (a) \textsf{NN} and (b) \textsf{SINN} for Twitter Abortion dataset.
% Words are highlighted according to attention scores.
}\label{fig:attention} 
\vspace{-6.5mm}
\end{figure}

\cref{fig:attention} visualizes the most important words in the profile descriptions identified by the attention mechanisms in \textsf{NN} and \textsf{SINN}   
for Twitter Abortion dataset.
As shown in this figure, NN tends to focus on meaningless or general words (e.g., ’https’, ’co’, ’year’), while SINN selects words (e.g., ’Pro’(-Life/Choice), ’Freedom’, ’Liberty’) that are more relevant to the given topic (i.e., Abortion).
With the help of prior scientific knowledge, \textsf{SINN} can extract meaningful features from rich side information like textual descriptions to better predict opinion evolution. 
To save space, we only report the results from one dataset in \cref{fig:network_synthetic,fig:hist_real,fig:attention}.

\section{Conclusion and Future work}
In this paper, we tackle the problem of modeling and predicting opinion dynamics in social networks.  
We develop a novel method that integrates sociological and social psychological knowledge and a data-driven framework, 
which is referred to as \textsf{SINN} (Sociologically-Informed Neural Network).  
Our approach is the first attempt to introduce the idea of physics-informed neural networks into opinion dynamics modeling.  
Specifically, we first reformulate opinion dynamics models into ordinary differential equations (ODEs). 
For stochastic opinion dynamics models, we apply the reparameterization trick to enable end-to-end training. 
Then we approximate opinion values by a neural network and train the neural network approximation under the constraints of the ODE. 
The proposed framework is integrated with matrix factorization and a language model to incorporate rich side information (e.g., user profiles) and structural knowledge (e.g., cluster structure of the social interaction network). 
We conduct extensive experiments on three synthetic and three real-world datasets from social networks. 
The experimental results demonstrate that the proposed method outperforms six baselines in predicting opinion dynamics. 

For future work, we try to integrate an opinion mining method \cite{medhat2014sentiment} into our framework  
to automatically obtain opinion labels for social media posts.  
This avoids expensive annotations. % and time-consuming 
We also plan to explore other side information such as mass media contents (e.g., news articles) and social graphs (e.g., follow/following relationship).  
Furthermore, although this paper focuses on four representative opinion dynamics models, our framework can be generalized for any opinion dynamics model. 

%%%%%%%%%%%%%%%%
%%%%%%%%%%%%%%%%

%%
\bibliographystyle{ACM-Reference-Format}
\bibliography{KDD22} 

\appendix
%\clearpage
\section*{Supplemental Material}
\section{Proposed method}\label{sec:app_proposed}

\subsection{Language Model}\label{sec:app_language}
The language model $h(\cdot, \theta_h)$ takes as input profile description $d^u$, and infers the hidden user representation $\textbf{h}_u$.  
User profile description $d^u$ consists of a sequence of $|d^u|$ words $d^u=\{w_1^u,...,w_{|d^u|}^u\}$.  
Input sentence $d^u$ is first padded to a fixed length of $N$ words with NULL tokens $d^u=\{w_{1}^u,...,w_{N}^u\}$ and tokenized. 
We pick 25 as the max sentence length (i.e., profile description) $N$.  
Original sentences with over $N=25$ words are truncated at the end. 
This truncation would not significantly affect the final result,  
because the most of profile descriptions contain no more than 25 words. 
91\% of users in the Twitter BLM dataset have less than 25 words ($|d^u| < 25$) in their profile descriptions; 
likewise 89\% of users in the Twitter Abortion dataset.  
In this work, we select a pre-trained BERT Transformer \cite{devlin2018bert} as the language model. 
The input tokens $d^u=\{w_{1}^u,...,w_{N}^u\}$ are mapped into a sequence of $b$-dimensional word vectors $\{{\bf w}_{1}^u,...,{\bf w}_{N}^u\}$ 
through the pre-trained BERT model. 
We add an attention layer on top of the last hidden layer of the frozen pre-trained BERT to learn the importance of each word in the sentence.  
Our attention layer takes as input the word vectors ${\bf w}_{n}^u$, 
and calculates the attention score for each vector: $e_n = {\bf w}_n^{\top} {\bf E}$, 
where $\textbf{E}\in\mathbb{R}^b$ is the context vector.  
The corresponding attention weight is defined as $\tilde{e}_n = \exp{(e_n)} / \sum_{n'}\exp{(e_{n'})}$. 
To compute the hidden representation of user $u$, we compute the weighted average of the word embeddings of each word in the profile description:  
$\textbf{h}_u=\sum_{n=1}^{N} \tilde{e}_n {\bf w}_{n}^u$.  

\subsection{Optimization}\label{sec:app_opt}
We solve the following minimization problem to find optimal parameters of neural network $\theta_f$, 
language model $\theta_h$ and the ODEs $\Lambda$: 
{\small\begin{align}
\theta_f^{\ast}, \theta_h^{\ast}, \Lambda^{\ast} = \argmin_{\{\theta_f, \theta_h, {\bf \Lambda}\}} \,\,\, 
\mathcal{L}(\theta_f, \theta_h, \Lambda; \mathcal{H}, \mathcal{D}),  
\end{align}}
\hspace{-1mm}where $\{\theta_f^{\ast}, \theta_h^{\ast}, \Lambda^{\ast}\}$ denote the optimal set of parameters. 
The loss function can be minimized by using a backpropagation algorithm.  
For the stochastic bounded confidence model (SBCM), during the backward pass, we can obtain the gradient of the discrete sample 
by computing the gradient of our continuous approximation $\tilde{\bf z}_{u}^t$ in \cref{eq:haty_u}.  
This allows the model to be optimized in an end-to-end manner. 

\subsection{Preidiction}\label{sec:pred}
During the test phase, we use the trained neural network to predict the future opinion $y^{\ast}$ of user $u^{\ast}$ at time $t^{\ast}$.  
The opinion can be predicted by calculating the hidden representation $\textbf{h}_{u^{\ast}}$ of user $u^{\ast}$, and  
feeding it into the trained neural network: $y^{\ast}=f(t^{\ast}, \textbf{e}_{u^{\ast}}, \textbf{h}_{u^{\ast}}; \theta_f^{\ast})$, 
where $\textbf{e}_{u^{\ast}}$ is the one-hot encoding of user $u^{\ast}$.

\section{Experiment}

\subsection{Datasets}\label{sec:datasets}
Experiments were conducted on three synthetic datasets and three real-world datasets.

\subsubsection{Synthetic dataset}\label{sec:synthetic_dataset}
We generated the syntetic datasets using the SBCM (stochastic opinion dynamics model) in \cref{eq:hegselmann,eq:p_uv}. 
We consider a social network with a set of $U=200$ users.  
For each dataset, the model was run for $T=200$ consecutive timesteps, resulting in a total of $I=40,000$ data points.  
At the first timestamp, initial opinions are randomly drawn from a uniform distribution between 0 and 1. 
In each timestep from 2 to 200, we randomly choose 15 users who initiate interaction with other users. 
When individual $u$ initiates an interaction, a partner $v$ is selected from a set of $U=200$ users with probability $p_{uv}^t$ defined by \cref{eq:p_uv}.  
User $u$ then updates her opinion $x_u(t + 1)$ at time $t + 1$ by weighted averaging her own opinion
$x_u(t)$ and the chosen partner's opinion $x_v(t)$ at time $t$: 
\begin{align}
x_u(t+1) = x_u(t) + \mu x_v(t),  
\end{align}
where parameter $\mu$ indicates the strength of influence.  
We set $\mu=0.1$ in all the simulations. 
The three datasets differ in the exponent parameter $\rho$ that reach different final states:  
$\rho=-1.0$ (opinion consensus), $\rho=0.1$ (opinion clustering), and $\rho=1.0$ (opinion polarization). 
Finally, we divided the continuous opinion value into five classes according to the range 
as highly negative ($-1.0 \text{--} -0.6$), negative ($-0.6 \text{--} -0.2$), neutral ($-0.2 \text{--} 0.2$), 
positive ($0.2 \text{--} 0.6$), and highly positive ($0.6\text{--}1.0$) and used these class labels as the input of the models.

\subsubsection{Twitter datasets}\label{sec:twitter}
We used Twitter API\footnoteref{note:twitter} to collect English tweets posted by Twitter users, specific to BLM and Abortion.  
The replies and retweets are not included in these datasets. 
All the tweets were pre-processed before annotation. 
First, we excluded the URLs, usernames (@user), and email addresses from the original tweets. 
To remove corporate accounts, bot accounts, and spammers, we excluded users 
who tweeted more than 30,000 tweets in total and less than three tweets,  
and those that contain the words ``bot'' and ``news'' in their username.  
To filter news tweets, we also excluded the tweet that contained specific keywords 
(i.e., 'news', 'call', 'tell', 'say', 'announce', 'state', 'country', 'city', 'council'). 
Moreover, we eliminated duplicate tweets. 
Annotators classified them into one of the following opinion classes: highly negative, negative, neutral, positive, highly positive, and not applicable.  
The ``not applicable'' label means the annotator could not make any judgment.  
Tweets with ``not applicable'' label were removed from the dataset. 
We also eliminated the tweets with ``neutral'' label since many of them contain only irrelevant information. 
The human-annotated labels are considered as the ground truth and the performance of the models is calculated with respect to these labels. 

Twitter profile descriptions are preprocessed as follows: 
(i) removing URLs, usernames and email addresses, 
(ii) eliminating stop words (``the'', ``is'', ``an'' etc.), 
(iii) remove extra white space, special characters, 
(iv) lowercasing, and 
(v) tokenization. 
We employed the BERT basic tokenizer (See \cref{sec:impl_detail}). 
\subsubsection{Reddit dataset}
The Reddit dataset was drawn from social media platform Reddit.  Using the Reddit API\footnoteref{note:reddit}, we collected a total of 70,876 Reddit posts 
between April 30th and November 3rd, 2020. 
We focus on two major communities (subreddits) discussing conservative and libertarian politics: 
\texttt{r/conservative} and \texttt{r/libertarian}.  
The dataset included anonymous user IDs and subreddits, as well as timestamps. We preserved only those users who had between 5 and 1,000 posts, which resulted in a set of 1,335 users. 
We assign label 0 (negative to conservative) to each post if it belongs to \texttt{r/libertarian}, label 1 (positive) if it belongs to \texttt{r/conservative}.

\subsection{Comparison methods}\label{sec:baseline_settings}
In this subsection, we describe the settings for the baselines. 

Some existing methods (i.e., \textsf{Voter} and \textsf{AsLM}) are designed for time series data observed at regular time intervals.  
To compare these methods, we treat the most recent opinion value as the one at the previous time step. 

Moreover, as the four baselines of \textsf{DeGroot}, \textsf{AsLM}, \textsf{SLANT}, and \textsf{SLANT+} can handle only continuous opinion values rather than discrete class labels,  
we transform discrete opinion labels to continuous values in $[-1,1]$ using linear scaling. 
At the time of evaluation, we rescale the predicted opinion values back to the discrete opinion labels,  
by grouping them into sets of discrete labels according to the range 
as highly negative ($-1.0 \text{--} -0.6$), negative ($-0.6 \text{--} -0.2$), neutral ($-0.2 \text{--} 0.2$), 
positive ($0.2 \text{--} 0.6$), and highly positive ($0.6\text{--}1.0$). 

Since \textsf{SLANT} and \textsf{SLANT+} are primarily intended for predicting the opinion of the next post,  
we predict the long-term evolution of individuals' opinions during the future time window by iteratively predicting the next post. 
The predicted opinion values are used as the input of the next time step. 
This procedure is repeated until time $T+\Delta T$.

\subsection{Implementation details}\label{sec:impl_detail}
All code was implemented using Python 3.9 and PyTorch \cite{paszke2017automatic}. 
We conducted all experiments on a machine with four 2.8GHz Intel Cores and 16GB memory.  
For the neural networks (i.e., \textsf{NN}, the FNN of our \textsf{SINN}), 
we used the same number of hidden units $N_u$ in all hidden layers.  
For the language model of \textsf{SINN}, we used the tokenizer and pre-trained BERT from the Python library 
\texttt{pytorch-pretrained-bert}\footnote{https://github.com/huggingface/pytorch-pretrainedBERT}.  
The hidden size of $\text{BERT}_\text{BASE–uncased}$ is set to 768. 
The input dimension of the attention layer was 768 and the output dimension was equal to the number of hidden units $N_u$ of the FNN $f(\cdot)$.  
The model parameters were trained using the ADAM optimizer \cite{kingma2014adam} 
with $\beta_1=0.9$, $\beta_2=0.999$ and a learning rate of 0.001. 
By default, we set 128 as the mini-batch size, 1000 as the number of epochs, and $J=1$ as the number of collocation points. 

\begin{figure}[t]
    \centering
    \subfigure[Trade-off parameter $\alpha$]{\includegraphics[width=0.23\textwidth]{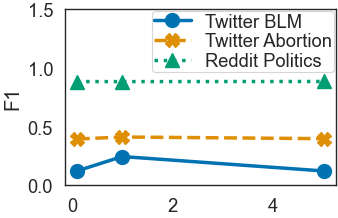}\label{fig:alpha}\hspace{1mm}} 
    %\subfigure[Trade-off parameter $\beta$]{\includegraphics[width=0.23\textwidth]{sensitivity_beta.png}\label{fig:beta}\hspace{1mm}} 
    \subfigure[Dimension of latent space $K$]{\includegraphics[width=0.23\textwidth]{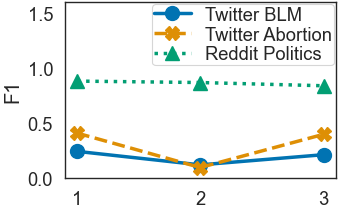}\label{fig:latent_dimension}} 
    \subfigure[number of layers $l$]{\includegraphics[width=0.23\textwidth]{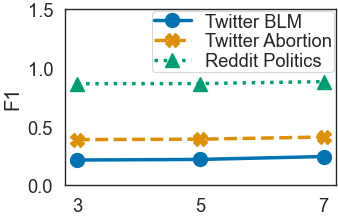}\label{fig:size_layer}\hspace{1mm}} 
    \subfigure[Number of units $N_u$]{\includegraphics[width=0.23\textwidth]{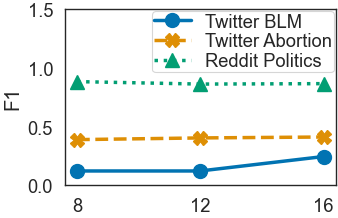}\label{fig:size_unit}} 
    \vspace{-3mm}
    \caption{Prediction performance with different values of hyperparameters using F1 score. }
    \label{fig:sensitivity_nn}
\vspace{-3mm}
\end{figure}

\begin{figure}[t]
    \centering
    \subfigure[Synthetic datasets]{\includegraphics[width=0.23\textwidth]{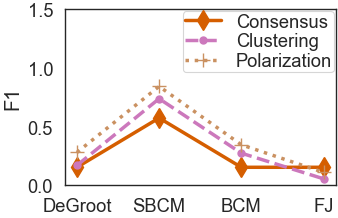}\label{fig:odm_synthetic}\hspace{1mm}} 
    \subfigure[Real-world datasets]{\includegraphics[width=0.23\textwidth]{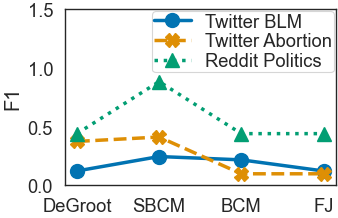}\label{fig:odm_real}} 
        \vspace{-3mm}
    \caption{Prediction performance with different opinion dynamics models using F1 score. }
    \label{fig:sensitivity_odm}
\vspace{-3mm}
\end{figure}

\subsection{Additional Results}
\subsubsection{Sensitivity Analysis}\label{sec:sensitivity}
We investigate the sensitivity of \textsf{SINN} on the core hyperparameters.  
Due to space limitations, we present only the results on the real-world datasets in \cref{fig:sensitivity_nn}.  

%\textbf{Trade-off Hyperparameters $\alpha$ and $\beta$. }
%As the loss function $\mathcal{L}$ of our proposed method involves multiple components, 
%we conduct several ablation experiments to investigate their individual contributions to the prediction performance.
\cref{fig:alpha} shows the prediction performance (in terms of F1) by varying the trade-off hyperparameter $\alpha$ in \cref{eq:totalloss}. 
\textsf{SINN} achieves the best performance when $\alpha=1.0$ for all the real-world datasets.  

In \cref{fig:latent_dimension}, we vary the dimension of latent space $K$. 
Our \textsf{SINN} gives the best results when $K=1$ for Twitter BLM dataset and Reddit Politics dataset; and $K=3$ for Twitter Abortion dataset.  
% Further results on parameter study 

%\textbf{Neural Network Architecture. }
%Here we examine the effects of the hyperparameters for the neural network of \textsf{SINN}. 
In \cref{fig:size_layer,fig:size_unit}, we show the impact of neural network architecture 
by varying the number of layers $L$ and units $N_u$ in the neural network.  
%Based on the results in \cref{fig:ablation_nn}, we set the number of layers $L$ as 16 and the number of units $N_u$ as 7 for all experiments.  
We can observe that \textsf{SINN} yields robust performance with respect to the size of the neural network 
(i.e., the number of layers $L$ and the number of units per layer $N_u$). 

We also evaluate the importance of side information (i.e., user profiles) by comparing \textsf{SINN} with and without profile descriptions.
For Twitter Abortion dataset, the use of such information improves the F1 score by 10.2\%. 
Meanwhile, it does not show improvement in prediction accuracy for Twitter BLM dataset.  
In future work, we will explore different choices of the language model. 

\subsubsection{Impact of Opinion Dynamics Models}\label{sec:choice_odm}
We evaluate different choices of opinion dynamics models on prediction performance.    
% We evaluate different design choices of our model on classification accuracy. 
\cref{fig:sensitivity_odm} reports the F1 results of \textsf{SINN} with four different opinion dynamics models: DeGroot model, FJ model, 
bounded confidence model (BCM), stochastic opinion dynamics model (SBCM). 
%We report the F1 results on the synthetic datasets in \cref{fig:ablation_odm}. 
For all the synthetic datasets, the SBCM outperforms all other opinion dynamics models explored in this study. 
That is, \textsf{SINN} successfully identifies the original opinion dynamics model that generated the synthetic datasets (i.e., SBCM) from a set of candidates. 
\textsf{SINN} also gives the best results on the SBCM for the real-world datasets.
It suggests the importance of considering the stochastic nature of social interaction.

\end{document}